\documentclass{mem}
\usepackage{natbib}
\usepackage{txfonts}
\usepackage{balance}
\usepackage{graphicx}
\usepackage[a4paper,breaklinks,dvipdfm]{hyperref}
\idline{75}{282}
\begin{document}

\title{Massive black holes interactions during the assembly
of heavy sub-structures in the centre of galaxy clusters}

   \subtitle{}

\author{
M. \, Donnari \inst{1,2} 
, \, M. \, Arca-Sedda \inst{1}
\and M. \, Merafina \inst{1}
          }

\institute{
Department of Physics, Sapienza University of Rome\\
Piazzale Aldo Moro 5, I-00185 Rome, Italy
\and
Department of Physics, Tor Vergata University of Rome\\
Via O. Raimondo 18, I-00173 Rome, Italy \\
\email{martina.donnari@uniroma1.it}
}

\authorrunning{Donnari }

\titlerunning{Galaxy merging in galaxy clusters}

\abstract{
We performed a series of direct N-body simulations with the aim to follow the dynamical evolution of a galaxy cluster (GC) ($M_{clus}\simeq 10^{14} M_{\odot}$) in different environment. The results show the formation of heavy sub-structures in the cluster centre in consequence of multiple merging among the innermost galaxies.
Moreover we investigate the dynamics of super-massive black holes (SMBHs) residing in the centre of galaxies that form the most massive sub-structure.

}
\maketitle{}

\section{Introduction and numerical method}
The innermost regions of galaxy cluster are good laboratories to study merging among galaxies that move, due to dynamical friction, within their host cluster.
In this way they can reach GC core and they can collide and merge each other, driving eventually the formation of a central elliptical galaxy through galactic cannibalism \citep{1977ostriker}.
Using a modified version of a direct N-body code \texttt{HiGPUs} \citep{2013spera}, we simulated a galaxy cluster with a mass $M_{clu} \simeq 10^{14} M_{\odot}$ composed of 240 galaxies, modelled according to a Dehnen profile and distribuited as a King profile. We performed several simulation studying the dynamical evolution of GC in presence of external potentials, investigating how the presence of different environments can modify the structural properties of GC \citep{2016Donnari}. 
The action of dynamical friction (df) \citep{2014Mas} leads to a concentrations of heavy galaxies in the GC centre thus allow them to easly merge.
We found that a central massive structures (MCS) forms, in all models, over a time-scale of about 3 Gyr.
Although the assumption that the majority of galaxies host at their centres SMBHs is widely accepted, it is not well clear how they interact during multiple galaxy mergings.
Because of this, with a modified version of a direct N-body code that include the algorithmic regularization (AR) scheme, we investigate the evolution of SMBHs contained within the merging galaxies. 

\begin{figure*}[t!]
\resizebox{\hsize}{!}{\includegraphics[clip=true]{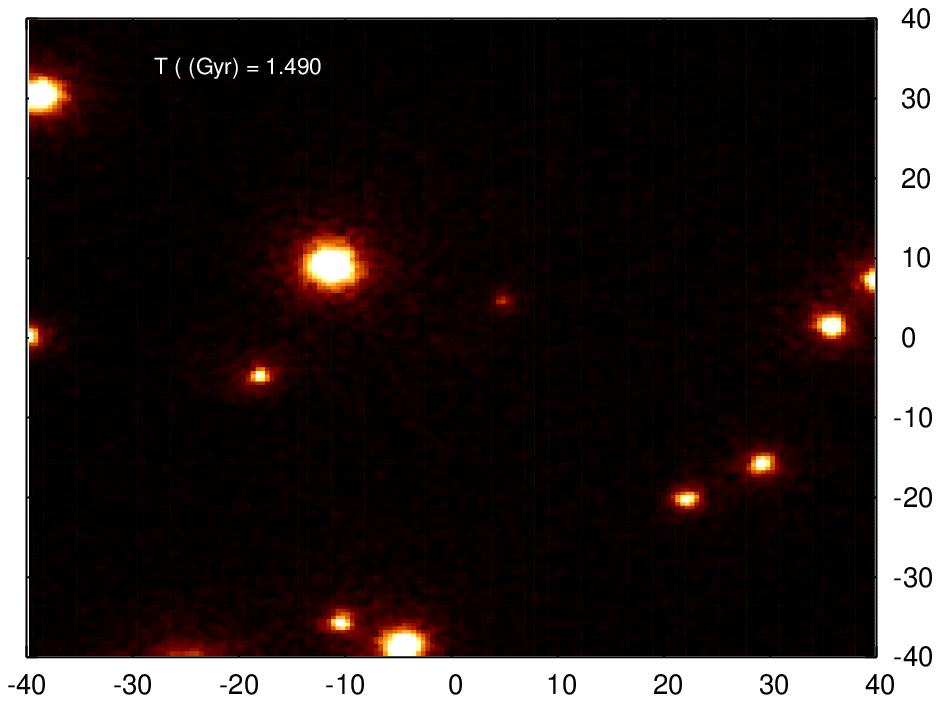}\includegraphics[clip=true]{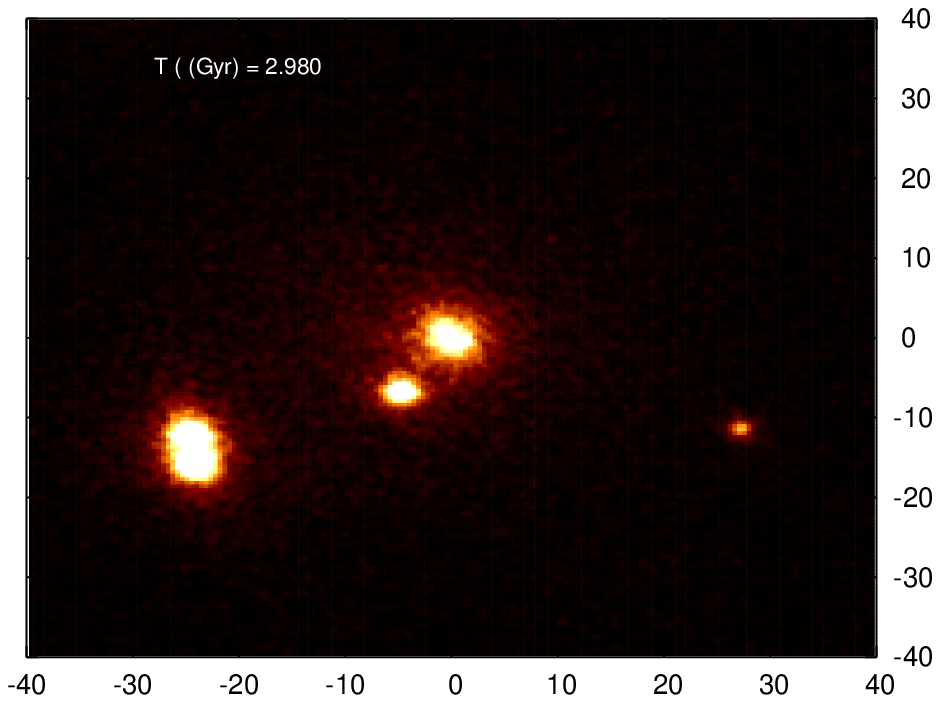}\includegraphics[clip=true]{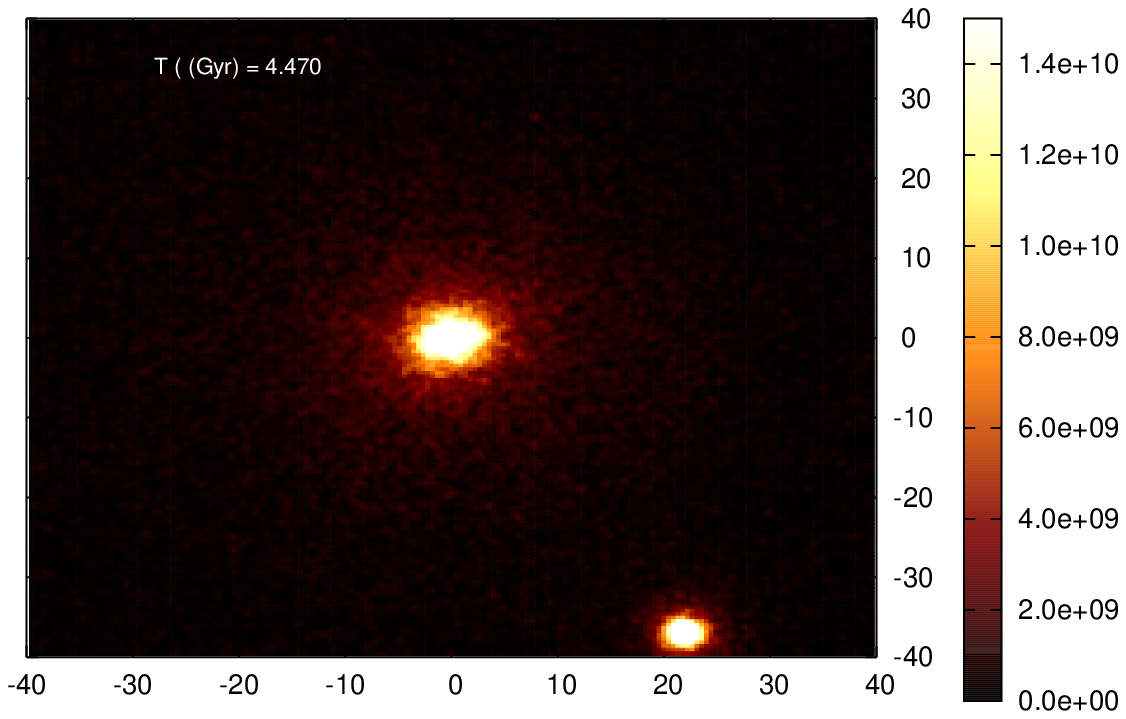}}
\caption{
\footnotesize
Some snapshots of the massive central system assembly.}
\label{snap}
\end{figure*}

\begin{figure}[]
\resizebox{\hsize}{!}{\includegraphics[clip=true]{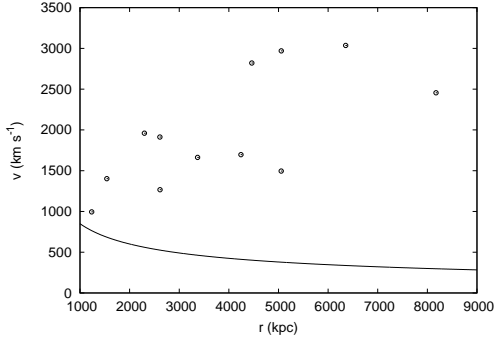}}
\caption{
\footnotesize
SMBHs velocities as a function of distance to the MCS centre. Open circles are the escaping SMBHs, the curve represents the MCS escape velocity.}
\label{vel}
\end{figure}

\begin{figure}[]
\resizebox{\hsize}{!}{\includegraphics[clip=true]{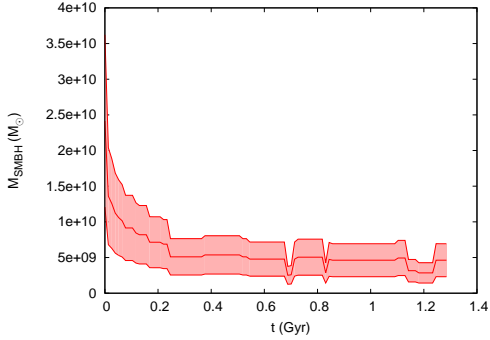}}
\caption{
\footnotesize
Mass of the SMBHs that move within $10$ kpc from the centre of the central structure.}
\label{BH}
\end{figure}
\section{Results and conclusions}

Our outcomes show that 21 galaxies reach the GC core within 5 Gyr, driving the formation of a sub-structure (Fig. \ref{snap}) with a mass $M_{MCS}=7\times 10^{12} M_{\odot}$, nearly 8$\%$ of the mass of the whole GC, quite close to the observed values.
In order to investigate the evolution of a population of SMBHs, we assume that each merging galaxy can hosts a SMBH in its centre. Their masses are assigned by using the relation $M_{SMBH}\simeq 2.4 \times10^{-3} M_{gal}$.
Since the AR scheme is required only when the SMBHs are at distance of 10-100 pc, we simulated their evolution using a semi-analytical model discussed in \cite{2014Mas} and \cite{2014Manuel}, over a time scale of 7.8 Gyr.
Results show that 12 SMBHs leave the system with a velocity $v=1000\div 3000$ km/s \citep{2014Miki} (Fig. \ref{vel}) . 
On the other hand, 4 SMBHs evolve, collide and merge, leading to the formation of a final hole with mass $M_{BH}^{final}=5\times10^9 M_{\odot}$ (Fig. \ref{BH}).
It is important to stress that this result is around $1.5 \times 10^{-3}$ times the host mass, close to the observed correlation between SMBHs and their host masses, suggesting that this relation can be regulated by galaxy merging and SMBH interactions.

\bibliographystyle{aa}

\end{document}